\documentstyle[12pt,epsf]{article}

\newcommand{\bq}{\begin{equation}}
\newcommand{\fq}{\end{equation}}
\newcommand{\bqr}{\begin{eqnarray}}
\newcommand{\fqr}{\end{eqnarray}}
\newcommand{\non}{\nonumber \\}

\newcommand{\xpv}[1]{\langle #1  \rangle}

\newcommand{\dspst}{\displaystyle}

\newcommand{\rf}[1]{(\ref{#1})}

\def\npb#1#2#3{Nucl. Phys. {\bf{B#1}} #3 (#2)}
\def\plb#1#2#3{Phys. Lett. {\bf{#1B}} #3 (#2)}

\def\prd#1#2#3{Phys. Rev. {D \bf{#1}} #3 (#2)}

\def\alp{\alpha}       \def\gam{\gamma}
\def\del{\delta}   \def\eps{\epsilon} 
        
    \def\kap{\kappa}   \def\lam{\lambda}
    
 \def\vphi{\varphi}

\def\Gam{\Gamma}   \def\Del{\Delta}   
     
\def\Ome{\Omega}

\def\npb#1#2#3{Nucl. Phys. {\bf{B#1}} (#2) #3}
\def\plb#1#2#3{Phys. Lett. {\bf{#1B}} (#2) #3}

\def\prd#1#2#3{Phys. Rev. {D \bf{#1}} (#2) #3}

 \def\cN{{\cal N}} \def\cO{{\cal O}}

\def\pa{\partial}

\def\rar{\rightarrow}

\def\one{1\!\!1\,\,}

\newcommand{\tr}{\mbox{Tr}}

\def\ove#1{\frac{1}{#1}}

\def\bk{{\vec{k}}}
\begin{document}

\thispagestyle{empty}

\begin{flushright}
\begin{tabular}{l}
ANL-HEP-PR-98-143 \\
ITP-SB-98-71 \\
\\
hep-th/9901144
\end{tabular}
\end{flushright}

\vspace{18mm}
\begin{center}

\baselineskip=20pt 
{\bf Holographic Normal Ordering \\ and \\ Multi-particle
States in the AdS/CFT Correspondence}
\baselineskip=16pt

\vspace{18mm}

{Gordon Chalmers${}^{\dagger}$ and Koenraad Schalm${}^*$
 }\footnote{E-mail addresses: chalmers@pcl9.hep.anl.gov and
konrad@insti.physics.sunysb.edu}
\\[10mm]
${}^{\dagger}${\em Argonne National Laboratory \\
High Energy Physics Division \\
9700 South Cass Avenue \\
Argonne, IL  60439 } \\ [5mm]
${}^*${\em Institute for Theoretical Physics  \\
State University of New York       \\
Stony Brook, NY 11794-3840 }

\vspace{20mm}

{\bf Abstract}

\end{center}

The general correlator of composite operators of N=4 supersymmetric
gauge field theory is divergent. We introduce a means for
renormalizing these correlators by adding a boundary theory on the AdS
space correcting for the divergences. Such renormalizations are not
equivalent to the standard normal ordering of current algebras in two
dimensions. The correlators contain contact terms that contribute to
the OPE; we relate them diagrammatically to correlation functions of
compound composite operators dual to multi-particle states.  \vfill

\setcounter{page}{0}
\newpage
\setcounter{footnote}{0}

\baselineskip=16pt

\section{Introduction}

Recently much progress has been made in the understanding of
superconformal $N=4$ super Yang-Mills theories through a holographic
description in terms of a IIB string (M-) theory on an anti-de Sitter
background \cite{mal, gub,  wit}. The low energy effective
supergravity of the string theory probes the strong coupling limit of
the CFT at large $N_c$; stringy/string-loop corrections 
correspond to the expansion in inverse powers of the 't Hooft
effective coupling $g^2_{YM}N_c$ and $N_c$. An explicit example is the
correspondence between type IIB supergravity on $ \mbox{AdS}_5 \times
S_5$ and $N=4$ super-Yang-Mills. Correlation functions in the latter 
at large $N_c$ may be determined from the supergravity theory 
through the relation:
\bq 
\prod_{j=1}^k \left.\left( {\delta\over
\delta\phi_{0,j}(\vec{z}_j)} \right) e^{iS_{\rm
sugra}[\phi(\phi_0)]}\right|_{\phi_{0,j}=0} = \langle \prod_{j=1}^k
{\cal O}^j(\vec{z}_j) \rangle_{CFT} \ .
\label{fundamental}
\fq
Here, $S_{sugra}$ is the bulk action of the supergravity theory
considered as a functional of the boundary values of the fields,
$\phi_{0,j}$, and ${\cal O}$ are composite (gauge
invariant) operators of the conformal Yang-Mills theory. These
operators are dual to the boundary values of the supergravity
fields in the sense that the latter act as sources
for the former.  In recent months CFT correlation functions have been 
analyzed using the holographic prescription in \cite{sfetsos}-\cite{freed2}.

In calculations exploring the correspondence, one item which has been
little addressed is the fact that the correlations of composite
operators are in general divergent. For example the Fourier transform
of  two-point functions of scalar operators ${\cal O}_{\Del}({\vec
x})$,  
\bqr   
\langle {\cal O}_{\Del}({\vec x}_1) {\cal
O}_{\Del}({\vec x}_2)\rangle &=&   {1\over \vert {\vec x}_1-{\vec
x}_2\vert^{2\Del}}  
~~~~\rar \non 
\langle {\cal O}_{\Del}({\vec k})
{\cal O}_{\Del}(-{\vec k})\rangle &=&
\frac{2^d\pi^{d/2}\Gam(d/2-\Del)}{\Gam(\Del)}
{\vec k}^{2(d/2-\Del)}\ , 
\fqr   
is ill-defined whenever the dimension $\Del$ of the operators is greater
than or equal to $d/2$. For a consistent description of the correspondence one
needs to provide a regularization of these short-distance
singularities to make the theory finite.  
This issue has been discussed from the CFT point of view in 
\cite{petkou1} and has been briefly mentioned in \cite{wit} in the context 
of the duality with AdS theories.  We  
introduce here a modification of the AdS/CFT correspondence 
through the addition of a boundary action, as a consequence of which the 
correlators are made finite. Different sets of boundary terms have been
considered in \cite{wit, sfetsos, tseyt1, muvis, arut}.

A regulatory scheme is to compute the supergravity Green's functions  at points
infinitesimally away from the AdS boundary. This IR cut-off for the gravity theory 
acts as an UV regulator for the CFT \cite{suswit,polpeet,kleb}. By introducing counterterms
with the associated scale we are thus in effect renormalizing the CFT
through the AdS boundary theory.  Of course this violates conformal
invariance. However, the prescription of \cite{gub,wit} considers not
the CFT as such but its perturbation by conformal operators, 
\bq
S_{N=4~SYM} ~\rar~ S_{N=4~SYM} + \int d^4\vec{x}
\phi_{0,j}(\vec{x})\cO^j(\vec{x}) \ ,  
\label{pert}
\fq 
where the source of the operator is the boundary value of a 
supergravity field. On the other hand conformal symmetry protects the dimensions 
of chiral primary operators
and their descendants. Thus for those operators, whose sources are elementary 
supergravity fields,  the introduction of a
regulator in intermediate steps of the calculation will not affect the
final answer provided one keeps the operator 
insertions at distinct points \cite{wit, muck, freed1}. However, 
correlations of multiple
operators may still diverge at short distances and these require the
introduction of counterterms \cite{joglekar}. This is in contrast to the 
finiteness of the unperturbed CFT. 

Insertions of operators in the SYM Green's functions would yield 
counterterms that are 
related to products of conformal operators at the same point.  These 
compound composite operators, e.g. $:{\rm Tr}F^2({\vec x}) 
{\rm Tr}F^2({\vec x}):$, are neither primary
nor descendants \cite{ferrara}.  As we will see such product
operators are dual to multi-particle supergravity states. Specific 
multi-particle states were found 
to be necessary in the AdS/CFT correspondence in \cite{jan,larsen}.  Here we 
propose that the coupling of such operators to the boundary values of 
supergravity fields is dictated by the above procedure. In \cite{liu} and 
\cite{freed2} there has been some speculation on
where such states might appear in exchange diagrams between
elementary supergravity fields.

In addition to the divergent nature of the correlation functions we 
find that explicit contact contributions appear in the evaluation 
of three- and four-point functions.  Their
Fourier transformed $k$-space expressions are divergent but they also
produce logarithms in the kinematic invariants; they contain cuts.  For 
example, the four-point function
contains a contact term of the following form 
\bq \xpv{\cO({\vec x}_1)\cO({\vec x}_2)\cO({\vec x}_3)\cO({\vec x}_4)} 
=  N^2 \delta^{(d)}({\vec x}_1-{\vec x}_2)  
\frac{1}{\vert{\vec x}_1-{\vec x}_3\vert^p}  
 \delta^{(d)}({\vec x}_3-{\vec x}_4) +{\rm permutations} \ , 
\fq  
together with products of multiple delta 
functions.  Through the delta functions this correlator resembles the {two}-point function of
the conformal operator $:\!\cO({\vec x})\cO(\vec{x})\!:$, dual to a
multi-particle supergravity state. The 
purpose of this work is to focus on coincident points, both how they
relate to the regulating of divergences and the appearance of contact terms 
in the calculation of correlation functions with physical implications. The 
results suggest a way to compute correlators of CFT compound composite 
operators from AdS supergravity.

The outline of this work is as follows.  In section 2 we recall how
conformal transformations constrain the form of correlators  at
distinct points.   In section 3 we review the Green's functions used
in the  computations and examine the asymptotic forms necessary for
the analysis of the contact-terms.  In Section 4 we examine the
divergences within the correlators by starting with a simple analysis
of the two-point functions. Our testing ground will be the
dilaton-axion sector of IIB supergravity on $\mbox{AdS}_5 \times
S_5$. In Sections 5 and 6 we do the same for three- and four-point
functions.  We discuss the multi-particle states in section 7. Contact 
term contributions are diagrammatically related to the multiparticle 
state correlators coming from bulk AdS multi-loop
supergravity.  Lastly, in section 8 we discuss implications and
extensions related to this work.

\section{Conformal Invariance Constraints} 
\setcounter{equation}{0}

In this section we briefly review the constraints imposed 
by conformal transformations on correlation functions of conformal operators.  

Conformal transformations preserve the line element up to a 
scale factor: 
\bq  
x_\mu\rightarrow x_\mu'(x) \qquad\quad 
\eta_{\mu\nu}dx^{\mu} dx^{\nu} = \Omega^{-2}(x) \eta_{\mu\nu} dx^\mu dx^\nu \ . 
\fq 
In $d$ dimensions they make up the 
conformal group $SO(d,2)$ generated by rotations, 
\bq  
x'_\mu = R_\mu{}^\nu x_\nu \qquad R_{\mu_1}{}^\nu R_{\mu_2\nu} 
 = \delta_{\mu_1\mu_2} \ ,~~ \Ome=1
\fq 
scale transformations, 
\bq  
x_\mu'= \lambda x_\mu \qquad \Omega^{-2}=\lambda^2 \ , 
\fq 
and special conformal transformations, 
\bq  
x_\mu'= \Omega^{-1}(x) \left[ x_\mu + v_\mu x^2\right] 
\qquad \Omega(x)= 1+2v\cdot x+ v^2x^2 \ . 
\fq 
Alternatively we may use inversion, 
\bq  
x_\mu'={x_\mu\over x^2} \qquad \Omega(x)=x^2 \ , 
\fq 
instead of the special conformal transformations to build 
up the generators of the group.

Conformal operators ${\cal O}^i$ of scale dimension  $\Delta$
transform as  
\bq   T\cdot {\cal O}^i = \Omega^{\Delta} D^i{}_j
\left[{\cal R}\right] {\cal O}^j \ .  \fq  This means that the
correlator of two such operators must behave under an   inversion as
\bq \xpv{ \cO_{\Del}(\vec{z_1})\cO_{\Del}(\vec{z_1})} \sim {1\over
\vert {\vec z}_1-{\vec z}_2\vert^{2\Delta}}  ~\rightarrow~ {1\over
\vert{\vec z}_1'-{\vec z}_2'\vert^{2\Delta}}  = \Omega^{\Delta}(z_1)
\Omega^{\Delta}(z_2) {1\over  \vert {\vec z}_1-{\vec
z}_2\vert^{2\Delta}} \ .  
\fq  
At coincident points, however, these
transformations are singular.  Conformal invariance constrains  the
correlators only when the points of the operators are at distinct
separated values.  Potential contact terms in the correlator are
allowed and probe the short-distance structure (UV region) of the conformal 
field theory.  As an example consider free-field
QCD Green's functions which are conformally invariant at
non-coincident points \cite{diff}.  Only when we investigate the
short-distance behaviour and compensate for infinities by the
introduction of counterterms do we find scale dependence. These 
short-distance singularities have been investigated in the context of 
conformal field theories in \cite{petkou1}. 

Note that
there are no UV non-renormalization theorems for correlations of
composite operators in $N=4$ super Yang-Mills theory, the reason being
that they are not finite. Non-renormalization theorems as proposed 
in \cite{freed1,min} refer only to the independence of correlators 
on the microscopic coupling $g^2_{YM}$.

\section{Asymptotic form of Greens functions}
\setcounter{equation}{0}

In this section we examine the relevant limits of the 
bulk-bulk and bulk-boundary kernels necessary for an exact 
evaluation of the contact contributions to the multi-point 
correlation functions examined in later sections. The theory we 
consider is the dilaton-axion sector of IIB supergravity on 
$\mbox{AdS}_5\times S_5$ with action
\bq
S=\frac{1}{2\kap^2}\int d^{5}x \sqrt{g(x)}(-R+12/A^2) + 
g^{\mu\nu}\left[\pa_{\mu}\phi \pa_\nu\phi + 
e^{2\phi}\pa_{\mu}C\pa_{\nu}C\right] 
\fq
Note that the interaction between the dilaton and axion 
contains derivatives. For the background metric on $\mbox{AdS}_5$ 
we will use the half-space Poincar\'{e}
metric 
\bq
ds^2 = \frac{A^2}{x_0^2}\left(dx_0^2 + dx^idx^j \eta_{ij}\right),~~
i=1,...,d
\label{met}
\fq
where $x_0 \geq 0$. We set the AdS radius $A^2$ to unity in the 
remainder and we will keep the dimension $d=4$ abstract whenever possible.  
The metric $\eta_{ij}$ is Minkowski with
mostly plus signature, and $\eta_{dd}= -1$.  

The bulk-bulk correlator
for massive scalars is given by
\bqr
\xpv{\vphi(x) \vphi(y)} &\equiv& G(x,y) \\ &=& (x_0y_0)^{d/2}
\int_0^{\infty} \lam d\lam \int \frac{d^dk}{(2\pi)^d} e^{i\vec{k}
\cdot
(\vec{x}-\vec{y})}\frac{J_{\nu}(\lam x_0)J_{\nu}(\lam
y_0)}{\lam^2+\vec{k}^2-i\eps}  ,
\label{BB}
\fqr
where $\nu =\sqrt{m^2+d^2/4} >0$ and $\vec{k} \cdot \vec{x} \equiv
\sum_{i=1}^d k_ix^i$.  We denote with ${\vec x}$ a (Minkowski) 
four-vector on the boundary of AdS.  The $i\epsilon$ prescription 
was provided in \cite{ustwo}.   As the dilaton and axion are massless 
we shall consider this case in the remainder of this work. 

The bulk-boundary kernel $\Delta({\vec x},y)$ is found by 
taking the small $y_0$ limit,
\bq
\triangle({\vec y},z) = \lim_{y_0\rightarrow 0}
{1\over y_0^{d/2+\nu-1}} \partial_{y_0} G(y,z)
\fq 
The extra factor of $y_0^{-\nu}$ corrects for the asymptotic behaviour 
of the Green's function. To evaluate this limit we need the asymptotic 
behaviour of the Bessel functions. At $z \rar 0$ they behave as  
\bq
J_\nu (z) = {1\over \Gamma(1+\nu)} \left({z\over 2}\right)^\nu +
\ldots
\qquad
K_\nu (z) = {\Gamma(\nu)\over 2} \left({2\over z}\right)^\nu + \ldots \ ,
\label{smallbes}
\fq
and at $z\rightarrow\infty$ we have,
\bq
J_\nu (z) = \left({1\over {2\pi z}}\right)^{\frac{1}{2}}
\cos(z+\pi/2) +
\ldots
\qquad K_\nu(z) = \left({\pi\over 2z}\right)^{1\over 2} e^{-z} + \ldots
\ .
\fq
Using (\ref{smallbes}) we find 
\bqr
\triangle({\vec y},z) &=& (z_0)^{d/2}
\int_0^{\infty} \lam d\lam \int \frac{d^dk}{(2\pi)^d} e^{i\vec{k}
\cdot
(\vec{z}-\vec{y})}\frac{2}{\Gam(\nu)}\left(\frac{\lam}{2}\right)^{\nu} 
\frac{J_{\nu}(\lam z_0)}{\lam^2+\vec{k}^2-i\eps} \\ &=& 
{2\over \Gamma(d/2)} \int {d^dk\over
(2\pi)^d}~
 \left( {\vert k\vert z_0 \over 2}\right)^{d/2} K_{\nu}
 (\vert k\vert z_0)~ e^{i{\vec k}\cdot ({\vec x}-{\vec y})} \ .
\label{bdybulk} 
\fqr 
This kernel satisfies the appropriate Dirichlet boundary conditions, as 
may be verified.  For the massless fields $\nu=d/2$ and in this case 
explicit integration over the Fourier modes ${\vec k}$ 
gives the position-space form, 
\bq  
\triangle({\vec y},z) = {\Gamma(d)\over \pi^{d/2} \Gamma(d/2)} 
 \left({z_0\over z_0^2 +  ({\vec y}-{\vec z})^2}\right)^d
\label{xtriangle}
\fq 
The Dirichlet conditions on the bulk-boundary kernel may also be 
verified using the distributional form above,
\bq
\lim_{z_0 \rar 0} \Del(\vec{y},z) = \del^d(\vec{y}-\vec{z}) \ ,
\fq
and a related identity is 
\bq  
\lim_{z_0\rightarrow 0} z_0 \partial_{z_0} \triangle(\vec{y},{ z}) 
 = 0 \ . 
\fq 

In the following we will also need the small $z_0$ limit of 
the derivative of the bulk-boundary kernel, 
\bq 
{\cal F}(\vec{y},\vec{z})=\lim_{z_0\rightarrow 0} {1\over z_0^{d-1}} 
\partial_{z_0} 
\triangle(\vec{y},{z}) \ , 
\label{derprop1}
\fq 
Straightforward 
differentiation of (\ref{bdybulk}) gives, 
\bq 
{\cal F}(\vec{y},\vec{z}) = \lim_{z_0\rightarrow 0} {\Gamma(d+1)\over 
\pi^{d/2} \Gamma(d/2)} 
\left\{ - {z_0^2\over \left[ z_0^2 + ({\vec y}-{\vec z})^2 \right]^{d+1} } 
 + {({\vec y}-{\vec z})^2 \over \left[ z_0^2 + ({\vec y}-{\vec z})^2 
\right]^{d+1}} \right\} \ .
\label{derprop}
\fq 
The functional form of the limit does not permit a naive interpretation 
as a distribution, and we need to include a divergent coefficient.  
The limits of (\ref{derprop}) are: 
\bq  
{\vec y}-{\vec z}\neq 0 \quad :\quad {\cal F}(\vec{y},\vec{z})= 
{\Gamma(d+1)\over \pi^{d/2} 
 \Gamma(d/2)} {1\over ({\vec y}-{\vec z})^{2d}} 
\ ,
\fq 
and 
\bq  
{\vec y}-{\vec z}=0 \quad : \quad {\cal F}(\vec{y},\vec{z})= 
-{\Gamma(d+1)\over \pi^{d/2} \Gamma(d/2)} {1\over y_0^{2d}} \ . 
\fq 
We define the $z_0\rightarrow 0$ form of (\ref{derprop}) to be 
\bq  
\lim_{z_0\rightarrow 0} {1\over z_0^{d-1}} \partial_{z_0} 
\triangle({\vec y},z) = {\Gamma(d+1)\over \pi^{d/2} \Gamma(d/2)} 
\left\{ {1\over ({\vec y}-{\vec z})^{2d}} - {1\over \mu^d} \delta 
({\vec y}-{\vec z}) \right\}  \ . 
\label{Ffunclim}
\fq  
The coefficient $\mu$ may be regarded as an infinitesimal
distance  from the boundary at $z_0=0$; this can be explicitly
verified by evaluating \rf{derprop1} using the momentum space
formulation of the bulk-boundary kernel.

\section{Two-point Functions}
\setcounter{equation}{0}

Conformal invariance fixes the form of the two-point function of
chiral primary operators with dimension $\Del$, the bosonic form 
of which is, 
\bq
\langle {\cal O}_{\Delta_1}({\vec z}_1) {\cal O}_{\Delta_2} ({\vec z}_2)
\rangle =  {\delta_{\Delta_1 \Delta_2} \over
|{\vec z}_1-{\vec z}_2|^{\Delta_1+\Delta_2}}
\ , 
\label{twopoint}
\fq 
provided the points are kept distinct.  Correlators of descendents 
easily follow.  This two-point function is 
computed through the holographic correspondence by solving
for the boundary-boundary kernel between two different points on
the boundary of the anti-de Sitter compactification and integrating over 
a two-point insertion in the bulk.  

The Fourier transform of the general two-point correlator (\ref{twopoint}) 
\bq
\langle {\cal O}_{\Del}({\vec k}) {\cal O}_{\Del}(-{\vec k})\rangle 
= \frac{2^d\pi^{d/2}\Gam(d/2-\Del)}{\Gam(\Del)}  
 {\vec k}^{2(d/2-\Del)} \ , 
\label{tut}
\fq 
is divergent and must be regularized. This is not an ad hoc
requirement, but follows directly from the free-field evaluation
of the two-point function of the operator $\tr \phi^i\phi^j$ 
in $N=4$ super Yang-Mills theory.  In this example the contribution is a
one-loop self-energy graph whose divergence equals that
of \rf{tut}.

The necessity of introducing a
regulating scale may be considered as a conformal anomaly 
\cite{gub} in the $N=4$ super Yang-Mills theory.  This resembles
the naive scale dependence in the tree-level bosonic propagator
$\langle X(z) X(v)\rangle \sim \ln(z-v)$ in open (or closed)
string theory. The
value of the composite two-point correlator at coincident
points is not determined through conformal invariance but suffers from
operator product ambiguities and associated divergences.  
In $N=4$ super Yang-Mills theory we find that there are
further modifications of higher-point functions.

We will add to the two-point correlator a counterterm
that eliminates the pathology at short distance.  In $x$-space
counterterms are known to be provided in the form of distributions
with support at coincident points.  Such 
regularizations and renormalizations have been extensively studied 
in the differential regularization approach \cite{diff}.  We shall 
modify the two-point function to the following form,
\bq
\langle {\cal O}_{\Delta}({\vec z}_1) {\cal O}_{\Delta} 
({\vec z}_2) \rangle
 = {1 \over |{\vec z}_1-{\vec z}_2|^{2\Delta}} 
 + \alpha \mu^{2\Del-d-2n} \Box^{n} \delta^{(d)} ({\vec z}_1-{\vec z}_2)
\ ,
\label{propren}
\fq where $n =\lceil \Delta-d/2 \rceil$ is the integer part of
$\Del-d/2$.  $\mu$ is a dimensionful regulating scale that permits the
counterterm to be built out of operators with well-defined
classical scaling dimensions: $\Box^n$, where $n$ is an integer.

The coefficient $\alpha$ is determined by enforcing finiteness on the
Fourier transformed two-point correlator, and is divergent. With the
addition of the counterterm and after dimensional continuation the
Fourier transform of (\ref{propren}) yields   
\bqr   
\langle {\cal
O}_{\Delta}({\vec k}_1){\cal O}_{\Delta}({\vec k}_2)  \rangle
&=&(2\pi)^d
\del^{(d)}(\bk_1+\bk_2)\frac{2^d\pi^{d/2}}{\Gam(\Del)}\frac{1}{ { k}^{-2n}}\left({\gamma_1} + \gamma_2 
\ln ({\vec k}^2/\mu^2) \right)  \ .  
\fqr 
and is finite. The above follows from
minimal subtraction with $\alp \sim 1/(\Del-d/2-n)$.  All of the
two-point functions may be regularised in this manner.  The renormalization scale
$\mu$ may be thought of as an  infinitesimal distance from the
boundary of AdS.

Rather than modifying the correlation functions by adding contact
terms by hand, we add a boundary action with these counterterms to the
bulk anti-de Sitter supergravity.  Boundary term additions have
been considered before within the purely gravitational anti-de Sitter action,
with the addition of the ``Gibbons-Hawking'' term \cite{wit,muck,tseyt1}
and in the work of \cite{sfetsos,arut}.  In our case we include the boundary
term, 
\bq 
S_{b.t.} = {\alpha\over 2} \mu^{2\Del-d-2n} \int d^dx ~
\phi_0(x) \Box^{n} \phi_0(x) \ ,
\label{btterm}
\fq where the boundary values of the supergravity fields,
$\phi_0({\vec x})$ are dual to the composite fields ${\cal O}({\vec
x})$.  The functional variation of (\ref{btterm}),  
\bq
{\delta\over\delta\phi_0({\vec z}_1)} {\delta\over\delta\phi_0({\vec
z}_2)}  S_{b.t.} =  \alpha \mu^{2\Del-d-2n} \int d^d{\vec x}
~\delta^{(d)}({\vec z}_1-{\vec x})  \Box_x^{n} \delta^{(d)}({\vec
z}_2-{\vec x}) \ ,  
\fq  
reproduces the counterterm in
(\ref{propren}). This is exactly the procedure of composite operator
renormalization in the CFT. In this case, since the two-point function
is given to all orders by its free-field value, this can be
explicitly verified. The perturbed CFT, \bq S_{N=4~SYM} ~\rar~
S_{N=4~SYM} + \int d^4\vec{x} J_j(\vec{x})\cO^j(\vec{x}) \ ,   \fq
requires counterterms consisting of all operators $\cO$ of similar
dimension or less. In particular graphs with no external fundamental
fields will require counterterms formed by products $J^n(\vec{x})\cdot
\one$. In the AdS/CFT correspondence the sources $J(\vec{x})$ are the boundary
values of supergravity fields.

For two-point functions of correlators of operators corresponding to the 
dilaton and the axion we have $\Del=d$ and $n=\lceil d/2 \rceil$. In this 
case the constants $\alpha$ and $\gamma_1$ and $\gamma_2$ are explicitly 
\bqr 
\alp
=\frac{(-1)^n}{n!}\ove{d/2-n},  &{\dspst \gam_1 =
\frac{3(-1)^{n}}{2n!}}&, \gam_2 =\frac{(-1)^n}{n!} \ .
\label{valval}
\fqr    
Similar terms occur for the boundary values of all the other 
supergravity fields. 

\section{Three-point Functions}
\setcounter{equation}{0}

We next analyze how similar divergent behaviour of the three-point
functions can be compensated by adding additional interaction terms on
the boundary of the anti-de Sitter space.  We will examine
the correlator $\langle {\rm Tr} F  {\tilde F}({\vec x}_1) {\rm
Tr}F{\tilde F}({\vec x}_2) {\rm Tr}  F^2({\vec x}_3) \rangle$ whose
supergravity dual is the unamputated $\langle C({\vec x}_1) C({\vec
x}_2) \phi({\vec x}_3)\rangle$ amplitude.  It arises within the
AdS/CFT correspondence by considering the contributions from the
$CC\phi$ vertex, 
\bq 
S_{CC\phi} = {1\over 2\kappa^2} \int
d^{d+1}x ~\sqrt{g(x)} \phi g^{\mu\nu} \partial_\mu C \partial_\nu C \
, 
\fq 
where $g_{\mu\nu}(x)$ is the background anti-de Sitter metric,
eq. \rf{met}.  The value of the correlator is computed through the use
of the bulk-boundary kernel to be, 
\bq 
A_{CC\phi}({\vec x}_1, {\vec x}_2, {\vec x}_3) = 
\int d^{d+1}y \sqrt{g(y)} \triangle({\vec x}_3,y)
g^{\mu\nu} \partial_\mu^{(y)} \triangle({\vec x}_1,y)
\partial_\nu^{(y)} \triangle({\vec x}_2,y) \ , 
\fq 
which through a
partial integration may be expressed as 
\bq 
A_{CC\phi} ({\vec x}_1,
{\vec x}_2, {\vec x}_3) = {1\over 2} \int d^d{\vec y}~ {1\over
y_0^{d-1}} \Bigl[ \triangle_{(1} \partial_{y_0} \triangle_{2)}
\triangle_3 -\triangle_1 \triangle_2 \partial_{y_0} \triangle_3
\Bigr]_{y_0=0}^{y_0=\infty}
\label{threepointex}
\ .  
\fq 
Here the parenthesis, $(1 2)$, denote symmetrization, $A_{(1}
B_{2)} = A_1 B_2 + A_2 B_1$. The bulk term vanishes due to the fact
that the bulk-boundary kernel in \rf{bdybulk} solves Laplace's
equation. We have used the shorthand $\triangle_i= \triangle({\vec
x}_i, y)$.

Inserting the kernels and the limits from section 3 we obtain 
two types of contributions to the three-point function, $A=A^{(2)}+A^{(3)}$, 
distinguished by the number of delta functions present.  We set the 
gravitational coupling $\kappa$ to unity from here on. 
The form $A^{(2)}$ contributes at doubly coincident points ${\vec x}_i$,  
\bqr  
A_{CC\phi}^{(2)} &=& {\Gamma(d+1)\over \pi^{d/2}\Gamma(d)} 
\int d^d{\vec y} ~ {1\over ({\vec y}-{\vec z}_2)^{2d}} 
\delta^{(d)}({\vec y}-{\vec z}_1) \delta^{(d)}({\vec y}-{\vec z}_3) \nonumber
\\ && + 
{1\over ({\vec y}-{\vec z}_1)^{2d}} 
\delta^{(d)}({\vec y}-{\vec z}_2) \delta^{(d)}({\vec y}-{\vec z}_3) \nonumber
\\ && 
- {1\over ({\vec y}-{\vec z}_3)^{2d}} 
\delta^{(d)}({\vec y}-{\vec z}_1) \delta^{(d)}({\vec y}-{\vec z}_2) \ .
\label{threepointtwo} 
\fqr 
The second one, $A^{(3)}$, contributes at triply-coincident points, 
\bq  
A_{CC\phi}^{(3)} = -{1\over \mu^d} {\Gamma(d+1)\over \pi^{d/2}\Gamma(d)} 
\int d^d{\vec y} ~ 
\delta^{(d)}({\vec y}-{\vec z}_1) \delta^{(d)}({\vec y}-{\vec z}_2) 
  \delta^{(d)}({\vec y}-{\vec z}_3) \ .
\label{threepointthree} 
\fq  
The fact that this three-point function only contains contact terms 
is in accordance with the constraints of conformal invariance \cite{freed1, min}.

To have a well-defined Fourier transform for the expressions in
$A^{(2)}$ we add the boundary action 
\bq 
S^{(2)}_{bdy,\phi CC} =
\alpha \int d^d{\vec x}~ C({\vec x}) C({\vec x}) \Box^{k+2} \phi({\vec
x})+ C({\vec x}) \Box^{k+2} C({\vec x})\phi({\vec x}) +  \Box^{k+2}
C({\vec x}) C({\vec x}) \phi({\vec x}) \ .
\label{elimthreepoint}
\fq 
Setting $\alpha$ to the same value as in eq.\rf{valval} the
divergence in the three-point correlator $A_{CC\phi}^{(2)}$  is
nullified.  We may also include a counterterm of the form  
\bq
S^{(3)}_{bdy,\phi CC} = {\beta\over\mu^d}  
 \int d^d{\vec x}~ C({\vec x}) C({\vec x}) \phi({\vec x}) \ , 
\fq 
to eliminate the contribution in $A_{CC\phi}^{(3)}$, with 
$\beta$ determined from (\ref{threepointthree}).

\section{Four-point Functions}
\setcounter{equation}{0}

Four-point correlation functions have the new feature that there are
two types of holographic Feynman diagrams to analyze: the one built
from two three-point bulk vertices exchanging an intermediate
supergravity field and the contribution from a bulk four-point vertex.
Scalar exchange contributions to the first diagram have been analyzed
several times \cite{tseyt2, freed3, ustwo, freed2}, and are known to be
reducible to an effective four-point vertex plus total
derivatives in the bulk coordinate.  As we found in the previous
section these total derivative terms contribute as contact terms to
the correlator.  Some of these have physical significance and contain 
cuts in the kinematic invariants.

It will suffice to consider the $s$-channel scalar exchange 
contribution to the correlator of four axions, $\xpv{CCCC}$,
\bqr
A^s_{CCCC}({\vec x}_i) &=& \int d^{d+1}y \sqrt{g(y)} ~
 \int d^{d+1}z \sqrt{g(z)} ~
 \left[ g^{\mu\nu}(y) \partial_\mu^y \triangle_1 \partial_\nu^y \triangle_2
 \right] \non  && 
\times G(y,z) ~
 \left[ g^{\mu\nu}(z) \partial_\alpha^z \triangle_3 \partial_\beta^z
 \triangle_2 \right] \ ,
\label{fourintrep}
\fqr
Following the steps in \cite{tseyt2, freed3} we partially integrate
\rf{fourintrep} with respect to both the $y_0$ and
$z_0$ coordinates.  After partially integrating the $y_0$ 
coordinate symmetrically we obtain a bulk four-point vertex 
contribution
\begin{eqnarray}
\nonumber
A^{s,~bulk}_{CCCC} = 
\frac{1}{2}\int {d^{d+1}y}\sqrt{g(y)}\int
  {d^{d+1}z}\sqrt{g(z)} &&\!\! \left[
\Del_{1} \Del_{2}\left(\frac{1}{\sqrt{g}}
  \pa_{\mu}\sqrt{g}g^{\mu\nu} \pa_{\nu}G(y,z)\right)
\right. \\
&&\hspace{-1.5in}
\left. -\Del_{(1}\left(\frac{1}{\sqrt{g}}\pa_{\mu}
   \sqrt{g}g^{\mu\nu}\pa_{\nu}\Del_{2)}\right) G(y,z)\right]
\left[ g^{\alpha\beta}(z) \pa_{{\alpha}}\Del_3
\pa_{{\beta}}\Del_4\right] \ .
\label{bulk}
\end{eqnarray}
plus boundary terms.
The second bulk term vanishes through the field equation for
the massless scalar and the first reduces to an effective four-point 
vertex. This contribution will be cancelled by those from the $t$- and 
$u$-channel exchange diagrams \cite{freed3}. Partially integrating the 
$z_0$-coordinate there remain three types of boundary terms 
\bqr
A^{s,~bdy}_{CCCC} &=& M_1+M_2+M_3 \non 
M_1 &=& \left.{1\over 4} \int d^d{\vec z} d^d{\vec y} {1\over (y_0z_0)^{d-1}} 
 \partial_{y_0} \triangle_{(1} \triangle_{2)} G(y,z) 
 \partial_{z_0} \triangle_{(3} \triangle_{4)} 
~\right|_{y_0,z_0=0}^{y_0,z_0=\infty} \\ 
M_2 &=&\left. -{1\over 4} \int d^d{\vec z} d^d{\vec y} {1\over (y_0z_0)^{d-1}} 
  \triangle_{1} \triangle_{2} \partial_{y_0} G(y,z) 
 \partial_{z_0} \triangle_{(3} \triangle_{4)} 
~\right|_{y_0,z_0=0}^{y_0,z_0=\infty} +(12 \leftrightarrow 34) \\  
M_3 &=&\left. {1\over 4} \int d^d{\vec z} d^d{\vec y} {1\over (y_0z_0)^{d-1}} 
  \triangle_{1} \triangle_{2} \partial_{y_0} \partial_{z_0} G(y,z) 
 \triangle_{3} \triangle_{4}  ~\right|_{y_0,z_0=0}^{y_0,z_0=\infty}  \ ,
\fqr 
in addition to similar contributions from the $t$- and $u$-channel.

It is straightforward to see that the $y_0$, $z_0$ contributions 
at $\infty$ all vanish and the only surviving ones are 
at the $y_0=z_0=0$ boundary. Evaluating these limits with the aid of 
section 3 we find that for $M_1$ the lower limit also vanishes. 
For $M_2$ we have two different contributions with triply and quadruply 
coincident points respectively, 
\bqr
M_2^{(a)} &={\dspst -{\Gamma(d+1)\over 4\pi^{d/2} \Gamma(d/2)}}& 
\int d^d{\vec y} d^d{\vec z} \delta^{(d)}({\vec z}_1-{\vec y}) 
\delta^{(d)}({\vec z}_2-{\vec y})  \delta^{(d)}({\vec y}-{\vec z}) 
 \non 
&&\times ~
 \delta^{(d)}({\vec z}_4-{\vec z}) {1\over ({\vec z}_3-{\vec z})^{2d}} + 
(3 \leftrightarrow 4)\ ,
\fqr 
and 
\bqr  
M_2^{(b)}&={\dspst {1\over 4\mu^d} {\Gamma(d+1)\over \pi^{d/2} \Gamma(d/2)}}& 
\int d^d{\vec y} d^d{\vec z} \delta^{(d)}({\vec z}_1-{\vec y}) 
\delta^{(d)}({\vec z}_2-{\vec y})  \delta^{(d)}({\vec y}-{\vec z}) 
\non  &&\times ~
\delta^{(d)}({\vec z}_4-{\vec z}) \delta^{(d)}({\vec z}_3-{\vec z}) + 
(3 \leftrightarrow 4)\ .
\fqr 
plus  contributions from ${\vec z}_1,{\vec z}_2\leftrightarrow 
{\vec z}_3,{\vec z}_4$.  

The terms of $M_3$ differs from $M_2$ in the 
arguments of the delta functions.  It also produces functions 
contributing at triply and quadruply coincident points but at 
different pairs, 
\bqr 
M_3^{(a)} &={\dspst {\Gamma(d+1)\over 4\pi^{d/2} \Gamma(d/2)}}& 
\int d^d{\vec y} d^d{\vec z} \delta^{(d)}({\vec z}_1-{\vec y}) 
\delta^{(d)}({\vec z}_2-{\vec y})  \delta^{(d)}({\vec z}-{\vec z}_3) 
 \non && \times 
\delta^{(d)}({\vec z}_4-{\vec z}) {1\over ({\vec y}-{\vec z})^{2d}} \ ,
\fqr 
and 
\bqr  
M_3^{(b)}&={\dspst -{1\over 4\mu^d} {\Gamma(d+1)\over \pi^{d/2} \Gamma(d/2)} }&
\int d^d{\vec y} d^d{\vec z} \delta^{(d)}({\vec z}_1-{\vec y}) 
\delta^{(d)}({\vec z}_2-{\vec y})  \delta^{(d)}({\vec z}-{\vec z}_3) 
 \non &&\times 
\delta^{(d)}({\vec z}_4-{\vec z}) \delta^{(d)}({\vec y}-{\vec z}) \ .
\fqr
Using the shorthand
\bq 
\delta_{ij}\equiv \delta^{(d)}({\vec z}_i-{\vec z}_j) \ . 
\fq 
the final expression for the correlator yields,
\bqr 
M_2^{(a)} &=& -{\Gamma(d+1)\over 4\pi^{d/2} \Gamma(d/2)}  
\left\{ 2\delta_{12} (\del_{13}+\delta_{14}){1\over ({\vec z}_4- {\vec z}_3)^{2d}}
\right\}   
\\  
M_2^{(b)} &=&  {1\over4\mu^d} {\Gamma(d+1)\over \pi^{d/2} \Gamma(d/2)} 
\left\{ 4 \delta_{12} \delta_{14} \delta_{13} 
  \right\} \ ,
\fqr 
The expression for $M_3$ is contained in,  
\bqr  
M_3^{(a)} &=&
{\Gamma(d+1)\over 4\pi^{d/2} \Gamma(d/2)} \left\{ \delta_{12}\del_{34}
{1\over ({\vec z}_3- {\vec z}_1)^{2d}}   \right\} \ , 
\label{tyt}
\fqr  
and  
\bqr
M_3^{(b)} &=& - {1\over4\mu^d} {\Gamma(d+1)\over \pi^{d/2} \Gamma(d/2)}
\left\{ \delta_{12} \delta_{14} \delta_{13} \right\} \ , 
\fqr   
Using $\delta_{12} f(x_1) = \delta_{12} f(x_2)$ the result is
symmetric under  ${\vec z}_1\leftrightarrow {\vec z}_2$ and ${\vec
z}_3\leftrightarrow {\vec z}_4$, as is manifest in the original graph.
Finally we need to also include the $t$- and $u$- channel diagrams to
find full Bose symmetry.  The sum of terms does not cancel. 

Proceeding as before we remove the infinities by the introduction of a 
four-point contribution to the boundary AdS theory.  The pure
(divergent) contact contributions of $M_2^{(b)}$ and $M_3^{(b)}$ are
completely removed similar to the $A^{(3)}$ contribution in the
previous section.  Further modification of the boundary theory by finite 
terms may also modify their correlations in the AdS picture.  As for 
the other terms, their Fourier transforms contain imaginary parts 
indicating that they contribute physically. In particular, the
Fourier transform of the scalar exchange to the $\langle CCCC\rangle$ 
correlator \cite{ustwo} contains the $s$-channel cut,   
\bq
{\rm Im} A_{CCCC}^{\phi, s} = -{\pi\over \kappa^2} \delta^{(4)}({\vec k}_1+
{\vec k}_2+{\vec k}_3+{\vec k}_4) ~{1\over 8} ({\vec k}_1+{\vec k}_2)^2 \ ,  
\fq  
and arises solely from the contact terms in \rf{tyt} above.  This is equivalent 
to the result in \cite{ustwo} after simplification.

Though we have not computed beyond four-point functions it is clear
from the previous results that the pattern of contact terms persists
in higher order. 

\section{Multi-particle states}
\setcounter{equation}{0}

Composite operator insertions in field theory lead to additional UV
ambiguities in correlation functions;  their  renormalizations require
the addition of counterterms of products of composite operators to the
correlation-functions.   Recent studies
have focused on position space, where  for non-coincident points
conformal symmetry imposes tight restrictions on the correlator
expressions.  Operators at coincident points and counterterms have not
been considered in detail.  In this section  we discuss these
singularities based on the renormalization of composite operators.

The AdS/CFT prescription comes with a natural
regulator: the  infinitesimal distance from the boundary of AdS. We
have found that the holographic supergravity description of the CFT
suffers from similar divergences when one tries to take this
distance to zero.  Since one expects
the full string theory embedding to be finite one might ask how string
theory will account for these in the low-energy limit.  This will not come 
from including stringy effects. The free-field 
result for the two- and three-point functions are
conjectured to be exact to all orders in $g^2_{YM}N_c$ (stringy) and $N_c$
(string loop) corrections, yet the answer is still divergent.  It
could be that these divergences are an additional feature of string
theory in a D-brane (Ramond-Ramond)-background, that is not yet
understood. 

Considering $N=4$  super Yang-Mills theory as a
consistent theory by itself, one would like to correct for the
divergences by the introduction of counterterms.  We have corrected
for the divergences in the CFT correlation functions by the
introduction of a boundary action to the AdS bulk theory consisting of
a polynomial in the sources for the conformal operators.  Complete
composite operator renormalization in field theory also yields 
counterterms consisting
of compound composite operators which also come in a power series in the
source for the ``simple'' composite operator \cite{joglekar}.  In
double insertions of $\tr F\tilde{F}$ in Green's functions, for
instance, one would correct for the UV infinities by adding to the
$N=4$ SYM action a term 
\bqr
S &=& S_{N=4~SYM} + \int d^4\vec{x}~ J(\vec{x}) \tr F\tilde{F}(\vec{x}) 
 + S^{(counter)} \\
S^{(counter)} &=& \ldots
+ \frac{c_i}{\mu^{p-d}} \int d^d\vec{x}~ J(\vec{x})J(\vec{x}) 
 \cO_p^i(\vec{x})+ \ldots
\label{count}
\fqr 
where $\cO^i_p$ are operators having dimension $p$ ($p \leq 8$) 
consistent with the symmetries.  In particular the compound composite
$: \tr F\tilde{F}(\vec{x}) \tr F\tilde{F}(\vec{x}):$ is one of them,
though $N$-counting  arguments show that this term is suppressed as
$1/N$.  This term reflects on the supergravity side how the compound 
composite operator is dual to a multi-particle state.  

We therefore conjecture that such 
additional couplings should be included in the
AdS/CFT correspondence from the beginning (the renormalization breaks 
scale invariance).   This means we
should consider $N=4$ SYM theory plus the space of all deformations  
\bq
S^{N=4}_{CFT} = \ldots + \sum \mu_{j_1\ldots j_k} \int d^d{\vec x}
\phi_{0,j_1} \ldots \phi_{0,j_k}  {\cal O}_{j_1} \ldots {\cal O}_{j_k}
\ ,
\label{multi} 
\fq 
with the sources corresponding to boundary values of the AdS 
fields as dictated by the renormalization of Green's functions. 
The $\mu_{j_1\ldots j_k}$ are dimensionful coupling constants.  
Their scale depends on the renormalization scale $\mu$, the 
distance to the boundary, naturally.  Functionally differentiation 
of these operators in the theory gives rise to correlators of 
compound composites after one interprets the delta functions 
$\delta^{(d)}(0)= \mu^{-d}$.  Following the prescription in \cite{gub,wit} the
correlation functions of products of composite operators would then be
given by {\em loop}-like calculations in supergravity
(Point-splitting on the boundary indicates that these couplings would 
arise from coincident point limits of single-particle correlation functions 
and that the legs connecting to the boundary would be given by the 
bulk-boundary kernel).  Diagrammatically this is
understood from pinching boundary points of the external fields in
the original holographic Feynman diagram.  This loop-like picture describes 
how the multi-particle states interact with 
others, though such loops contribute at the same order as the regular 
tree-level diagrams after appropriate normalization \cite{adsdyn}.

Composite operator renormalization suggests these multiparticle
couplings as in \rf{count} and \rf{multi}.  At the same time we have physical contact term
contributions to the correlators of ``simple'' operators.  They
resemble diagrammatically the multiparticle
correlators; for instance, the two separate delta function contributions
$A_{CC\phi}^{(2)}$ and $A_{CC\phi}^{(3)}$ in  (\ref{threepointtwo})
and (\ref{threepointthree}) are pictorially identical to such one- and two-loop 
supergravity diagrams on AdS, respectively.  Our calculations in earlier sections 
indicate the presence in the OPE of 
contact terms which modify the expansion via  
\bq 
\cO_n({\vec x})\cO_n({\vec y}) = \sum_j 
 \frac{{\cal O}^{\Delta_j}}{({\vec z}-{\vec y})^{2n-\Delta_j}}
+\del^{(d)}({\vec x}-{\vec y}) \cO_{2n-d} +\ldots \ .
\label{jjj}
\fq 
Such contact contributions are usually required for consistency with the 
Ward-identies of the theory \cite{petkou1, sei2, green, kutasov} and are 
therefore not subject to renormalization ambiguities. 

Finally it is worth noting that the AdS/CFT correspondence requires the
existence of multi-particle states which occur in specific
long-multiplets of the AdS supergroup \cite{jan, larsen}. The above proposal  
provides a way of computing correlators involving such states.

\section{Conclusion} 
\setcounter{equation}{0}

We have provided a scheme for the renormalization of composite
operator correlation functions within the AdS/CFT correspondence in
the large $N$ limit  involving the addition of a boundary supergravity
theory to the  bulk gauged supergravity theory.  The general
correlation function is made finite by modifying the correspondence
as,  
\bq 
\prod_{j=1}^k \left.\left( {\delta\over
\delta\phi_{0,j}(\vec{z}_j)} \right) e^{iS_{\rm
sugra}[\phi(\phi_0)]+iS_{bt} }\right|_{\phi_{0,j}=0} = \langle
\prod_{j=1}^k {\cal O}^j(\vec{z}_j) \rangle_{CFT}^{ren} \ ,  
\fq
where order by order the boundary theory $S_{bt}$ is chosen  to
contain counterterms renormalizing the correlators, as calculated  in
this work.  This scheme does not maintain conformal invariance.  Multi-trace 
states, whose dimensions are not protected, appear at coincident points 
where ultraviolet singularities occur.  Although the {\em microscopic} 
$N=4$ super Yang-Mills theory is UV finite \cite{finite} there are no
non-renormalization  theorems for correlators of composite operators.
Straightforward calculations of the two-point and three-point
functions, for example, show the presence of divergences and the
need for a renormalization  scheme.  It would be interesting to find 
the renormalization in the string context. 

The correlators we have examined possess both contact terms in 
their explicit expressions and singularities which render their 
Fourier transform divergent.  The contact
terms in the form of delta function distributions contribute
at coincident points - they appear discontinously  in the
OPE.  Diagrammatically they resemble correlators involving multi-particle
states which are dual in the AdS/CFT correspondence to products of
single-trace operators ${\cal O} ({\vec x})$ at the same point. The 
dimensions of the latter are in general not protected which is related 
to the divergences in the theory which occur at short distances (when points 
collide).  The renormalization and effective theory description 
of the correspondence allows one to determine the couplings of such compound 
composite operators to supergravity fields.

Given these results it appears necessary to examine in greater detail the 
conformal structure of the correspondence in the field theory and possible 
deformations of the conformal theory with the operators discussed here.

\vskip .5in 

\subsection*{Acknowledgements}

We would like to thank Luis Alvarez-Gaum\'e, Geoff Bodwin, Jan de Boer, 
Nick Mavromatos, Martin Rocek, Duncan Sands, Samson 
Shatasvilli, Warren Siegel, George Sterman, Cosmas Zachos and 
especially Daniel Z. Freedman for discussions.   
G.C. would like to thank CERN for its hospitality and K.S. would like 
to thank Argonne National Laboratory during the final stages of this 
work.  The work of G.C. is supported in part by the U.S. Department of Energy, 
Division of High Energy Physics, Contract W-31-109-ENG-38.

\end{document}